\documentclass[11pt]{article}

\usepackage{amsmath}
\usepackage{graphicx}
\usepackage{amsfonts}
\usepackage{amssymb}
\usepackage{epsfig}
\usepackage{color}
\usepackage{psfrag}
\usepackage{epstopdf}

\newcommand{\EQ}{\begin{equation}}
\newcommand{\EN}{\end{equation}}
\newcommand{\be}{\begin{equation}}
\newcommand{\ee}{\end{equation}}
\newcommand{\bea}{\begin{eqnarray}}
\newcommand{\eea}{\end{eqnarray}}

\begin{document} \setcounter{page}{0}
\topmargin 0pt
\oddsidemargin 5mm
\renewcommand{\thefootnote}{\arabic{footnote}}
\newpage
\setcounter{page}{0}
\topmargin 0pt
\oddsidemargin 5mm
\renewcommand{\thefootnote}{\arabic{footnote}}
\newpage
\begin{titlepage}
\begin{flushright}
\end{flushright}
\vspace{0.5cm}
\begin{center}
{\large {\bf Structure of interfaces at phase coexistence. Theory and numerics}}\\
\vspace{1.8cm}
{\large Gesualdo Delfino$^{1,2}$, Walter Selke$^3$ and Alessio Squarcini$^{4,5}$}\\
\vspace{0.5cm}
{\em $^1$SISSA -- Via Bonomea 265, 34136 Trieste, Italy}\\
{\em $^2$INFN sezione di Trieste}\\
{\em $^3$Institute for Theoretical Physics, RWTH Aachen University, 52056 Aachen, Germany}\\
{\em $^4$Max-Planck-Institut f\"ur Intelligente Systeme, 
Heisenbergstr. 3, D-70569, Stuttgart, Germany}\\
{\em $^5$IV. Institut f\"ur Theoretische 
Physik, Universit\"at Stuttgart, Pfaffenwaldring 57, D-70569 
Stuttgart, Germany}\\

\end{center}
\vspace{1.2cm}

\renewcommand{\thefootnote}{\arabic{footnote}}
\setcounter{footnote}{0}

\begin{abstract}
\noindent
We compare results of the exact field theory of phase separation in two dimensions with Monte Carlo simulations for the $q$-state Potts model with boundary conditions producing an interfacial region separating two pure phases. We confirm in particular the theoretical predictions that below critical temperature the surplus of non-boundary colors appears in drops along a single interface, while for $q>4$ at critical temperature there is formation of two interfaces enclosing a macroscopic disordered layer. These qualitatively different structures of the interfacial region can be discriminated through a measurement at a single point for different system sizes. 
\end{abstract}
\end{titlepage}

\newpage
\section{Introduction}
\begin{figure}
\centering
\includegraphics[width=7cm]{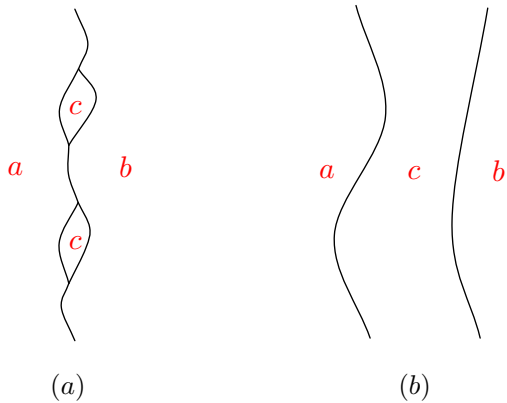}
\caption{Two different structures of the interfacial region: a third phase appears in bubbles along a single interface (a), or through a wetting layer between two interfaces (b).}
\label{wetting}
\end{figure}

In a statistical system at phase coexistence boundary conditions can be chosen to select a pure phase $a$ on the far left and a pure phase $b$ on the far right. The two phases are separated by an interfacial region that nearby a critical point will exhibit properties depending on the universality class of the system.
The determination of these properties is a classical problem of statistical mechanics \cite{Widom,Fisher}. Qualitatively, the simplest picture is that the interfacial region is produced by the fluctuations of an interface that sharply separates the two pure phases, but it is also natural to expect corrections associated to the notion of a {\it structure} of the interface. In a system admitting more than two phases, the first correction can be expected from drops of a third phase $c$ forming on the interface (Fig.~\ref{wetting}a). It is also possible that for different values of the parameters of the system the proliferation of these drops produces a macroscopic layer of the third phase in between two interfaces (Fig.~\ref{wetting}b). This second regime goes under the name of interfacial wetting (see \cite{Dietrich}). 

On the theoretical side, the two-dimensional case has emerged for the possibility of obtaining exact results for the universal interfacial properties. Initially, this was achieved for the Ising model, taking the scaling limit of lattice calculations \cite{Abraham}. This allowed, in particular, the determination of the magnetization profile across the interface, in the leading approximation of large system size that turns out to amount to sharp phase separation of pure phases. Only more recently it has been shown how exact results for the different universality classes can be obtained directly in the continuum limit, formulating the field theory of phase separation \cite{DV,bubbles}. This has allowed the determination of the corrections related to the structure of the interface \cite{DV}, of the quantitative properties of interfacial wetting \cite{bubbles}, of the long range correlations generated by the interface \cite{long_range}, as well as of effects of system geometry \cite{DS1,wedge,wedge2}. 

A simple criterion has also been deduced for interfacial wetting in two dimensions \cite{localization}. For the system with more than two coexisting phases, let $\lambda$ be the parameter that measures the distance from a second order phase transition point, so that the {\it bulk} correlation length diverges as 
\EQ
\xi\sim\lambda^{-\nu}\,, \hspace{1cm} \lambda\to 0\,.
\label{xi}
\EN
Then boundary conditions yielding interfacial wetting exist if $\nu\geq 1$. 

The exact field theory of phase separation also indicates the way of verifying the realization of interfacial wetting through a measurement at a single point. Consider the system on the strip depicted in Fig.~\ref{rectangle}, with boundary condition changing at the points $(x,y)=(0,\pm R/2)$, and the length scales satisfying
\EQ
R\gg\xi\gg a\,,
\label{lengths}
\EN
where $a$ is the microscopic scale (lattice spacing in a simulation). The first inequality is needed for phase separation to emerge, the second for the continuum description to apply. Then the excess probability of finding the third phase in the origin is suppressed as $R^{-1/2}$ in absence of interfacial wetting, and approaches a constant value for large $R$ in presence of interfacial wetting \cite{DV,bubbles}. More generally, the theory yields the order parameter profiles across the interfacial region and allows to deduce from them the interfacial properties (passage probabilities, interface structure). 

In this paper we consider the $q$-state Potts ferromagnet and compare the analytic predictions of the theory to the results of Monte Carlo simulations. The Potts model has been considered since the early studies of the global surplus of non-boundary states induced by an interface (interfacial adsorption) \cite{SP,SH}, and is a natural choice also for our present purposes. Indeed, in two dimensions the model is known to possess a phase transition which is of the second order up to $q=4$ and becomes of the first order for $q>4$ \cite{Baxter}. We will first consider the case $q=3$ below the critical temperature $T_c$, where the three ferromagnetic phases coexist. The relation (\ref{xi}) holds with $\lambda\sim T_c-T$ and $\nu=5/6$ \cite{Nienhuis}, so that we have a basic example for which no interfacial wetting is expected. On the other hand, for $q>4$ and $T=T_c$ the $q$ ferromagnetic phases coexist with the disordered phase, and the correlation length behaves as $\xi\sim e^{b^2/\sqrt{q-4}}$ for $q\to 4^+$, with $b=\pi$ for the square lattice \cite{BW}. The renormalization group arguments of \cite{DCq4} allow to associate this behavior to the relation (\ref{xi}) with $\lambda\sim q-4$ and $\nu=\infty$, so that interfacial wetting is expected. Strictly speaking the continuum limit is defined for $q\to 4^+$, but it was shown in \cite{DCq4} for bulk properties that field theory is able to provide a quantitative description as long as $\xi\gg a$. We will see here that this is true also for interfacial properties, comparing theory and numerical data at $q=10$, where $\xi$ is around ten lattice spacings. 

\section{Geometry and observables}
The $q$-state Potts model \cite{Wu} is defined on the lattice by the Hamiltonian
\begin{equation}
\label{01}
\mathcal{H} = -J \sum_{<i,j>} \delta_{s_{i},s_{j}} \, ,
\end{equation}
in which on each site the spin variable takes discrete values, often called colors, in the set $s_{i} \in \{ 1,\dots, q\}$, and the sum is restricted to nearest neighboring sites. Equal neighboring colors are favored by the ferromagnetic coupling $J>0$. 
The model is manifestly invariant under global permutations of the color variables. 

\begin{figure}
\centering
\includegraphics[width=11.5cm]{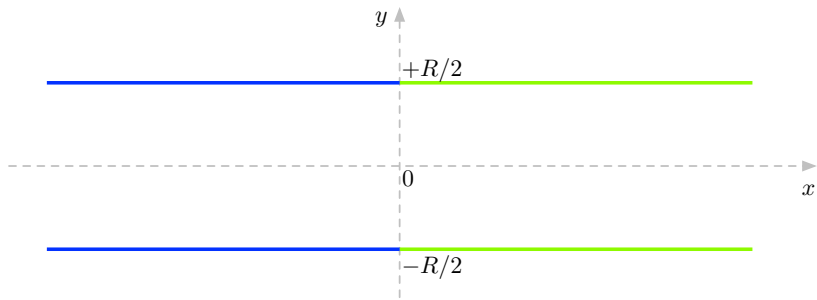}
\caption{Strip geometry considered in the paper. Along the edges of the strip the Potts spin variable is fixed to color $1$ for $x<0$ and to color $2$ for $x>0$.
}
\label{rectangle}
\end{figure}

For the study of phase separation we consider the model on the strip depicted in Fig.~\ref{rectangle}, with boundary spins fixed to color 1 for $x<0$ and to color 2 for $x>0$. Field theory yields the profiles for all the components of the order parameter and allows to deduce from them the properties of the interfacial region \cite{DV,bubbles}. For the purposes of this paper we focus on the probability 
\EQ
{\cal P}(x)=\sum_{c=3}^q\langle\delta_{s(x,0),c}\rangle=(q-2)\,\langle\delta_{s(x,0),3}\rangle
\label{profile}
\EN
of finding a non-boundary color (i.e. a color different from 1 and 2) at a point $x$ along the axis $y=0$; we adopted the notation $s(x,y)$ for the spin variable at the point of coordinates $(x,y)$. The second equality in (\ref{profile}) follows from the permutational symmetry of the Potts model, which also implies that ${\cal P}(x)$ is an even function of $x$. In particular we denote as
\EQ
B=\lim_{x\to\pm\infty}{\cal P}(x)
\label{asymp}
\EN
the asymptotic value in the limits in which the pure phases selected by the boundary conditions are recovered. 

The {\it bulk} correlation length $\xi$ that will appear in our subsequent formulae is the usual one defined by the large distance decay of the connected spin-spin correlation function 
\EQ
G(r)\sim e^{-r/\xi}\, 
\label{xi_decay}
\EN
in a pure phase at the given temperature. 

In the numerical simulations the infinitely long strip is replaced by a rectangle with $|x|<L/2$, $L\gtrsim R$, and $R$ satisfying (\ref{lengths}). Applying the Metropolis algorithm \cite{x}, we studied
lattices with\footnote{Throughout the paper lengths entering simulations are expressed in units of lattice spacing.} $L$ up to 402 and $R$ up to 251. Along the entire
boundaries the Potts spin variables were fixed, see Fig.~2. At given number
of colors, $q$, and given temperature, we typically averaged
over four to twelve realizations, using different random numbers. Usually,
for each realization, simulations with, at least, $10^7$ Monte Carlo
steps per site were performed. Similar choices had been tested and applied
before in a recent Monte Carlo study on interfacial properties
of two-dimensional Potts models \cite{y}. 

\begin{figure}
\centering
\includegraphics[width=9cm]{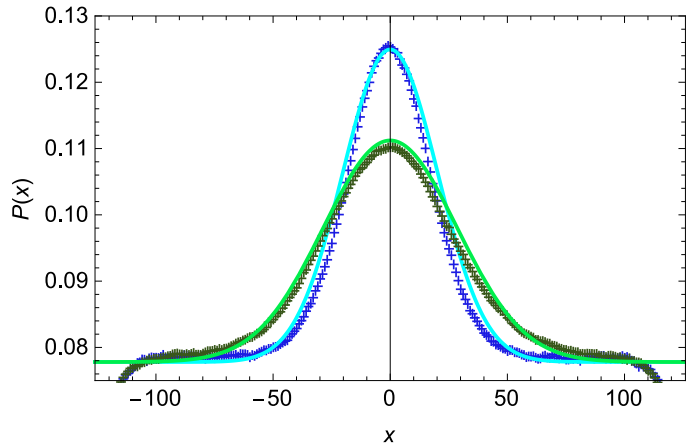}
\caption{Comparison of the analytic result (\ref{p3}) (solid curves) with the Monte Carlo data for $R=101$ (higher peak) and $R=201$ at $T=0.985\,T_c$. The numerical data are obtained for a system of horizontal size $L=252$. 
}
\label{figure03}
\end{figure}

\section{Single interface}
For $T<T_c$ the theory predicts a single interface (i.e. no interfacial wetting) for the $q$-state Potts model\footnote{The field theoretical derivation is performed for $q\leq 4$, but the main quantitative conclusions can be extended at least up to $q\approx 10$ following the arguments of \cite{bubbles,DCq4}.} \cite{DV,bubbles}. More precisely field theory yields the following result for the probability (\ref{profile}) of a non-boundary color along the axis $y=0$ \cite{DV}:
\EQ
\mathcal{P}(x) = B + (q-2)(q-1)\left(\frac{1}{q} - \frac{B}{q-2} \right) 2\xi\,C\,p_1(x)+\cdots\,,
\label{pq}
\EN
where 
\EQ
p_1(x)=\frac{\textrm{e}^{-\frac{x^{2}}{R\xi}}}{\sqrt{\pi R\xi}}\,;
\label{pd}
\EN
$p_1(x)dx$ is the probability that the interface intersects the axis $y=0$ in the interval $(x,x+dx)$, and then $\int_{-\infty}^\infty p_1(x)\textrm{d}x=1$. The dots in (\ref{pq}) correspond to subleading terms that are $O(1/R)$ in $x=0$. The result (\ref{pq}) has the simple interpretation that, at leading order for $R$ large, the appearance of the third phase is due to the formation along the interface of the droplets\footnote{The droplets should be associated to the ``irreducible pieces'' whose presence along the interface of the square lattice $q$-state Potts model below $T_c$ is rigorously established in \cite{CIV}.} depicted in Fig.~\ref{wetting}. This formation occurs with a probability proportional to the pure number $C$, which for $q\leq 4$ is characteristic of the universality class and can be determined exactly; its value is $0$ at $q=2$, $1/(2\sqrt{3})$ at $q=3$, and $2/(3\sqrt{3})$ at $q=4$ \cite{DV}. Hence we have in particular 
\begin{equation}
\mathcal{P}(x) = B + \left( \frac{2}{3} - 2B \right) \sqrt{\frac{\xi}{3\pi R}} \,\textrm{e}^{-\frac{x^{2}}{R\xi}}+\cdots \,, \hspace{1cm}q=3\,.
\label{p3}
\end{equation}
This result is compared in Fig.~\ref{figure03} with numerical determinations of the profile ${\mathcal P}(x)$ in the square lattice three-state Potts model at $T=0.985\, T_c$. As a matter of fact, ${\cal P}(x)$ is the only quantity we measured, extracting in particular the value $B\simeq 0.0778$ to be used in (\ref{p3}). We then obtained $\xi\simeq 8.1$ as the value to be used in (\ref{p3}) in order to best reproduce the numerical profiles in function of $R$. This result for $\xi$ is consistent with the estimate coming from the scaling formula $\xi\simeq\xi_0^\pm|T-T_c|^{-5/6}$ for $T\to T_c^\pm$, together with results of \cite{Jasnow,DC98}. Indeed, in \cite{Jasnow} the value $(\xi_0^+)_{2nd}\simeq 0.52$ was found for the critical amplitude of the {\it second moment} correlation length on the square lattice, while it is known from \cite{DC98} that $\xi_0^+\simeq(\xi_0^+)_{2nd}$ and that $\xi_0^+=2\xi_0^-$. This information produces the estimate $\xi\approx 8$ at $T=0.985\,T_c$. 

\begin{figure}
\centering
\includegraphics[width=10cm]{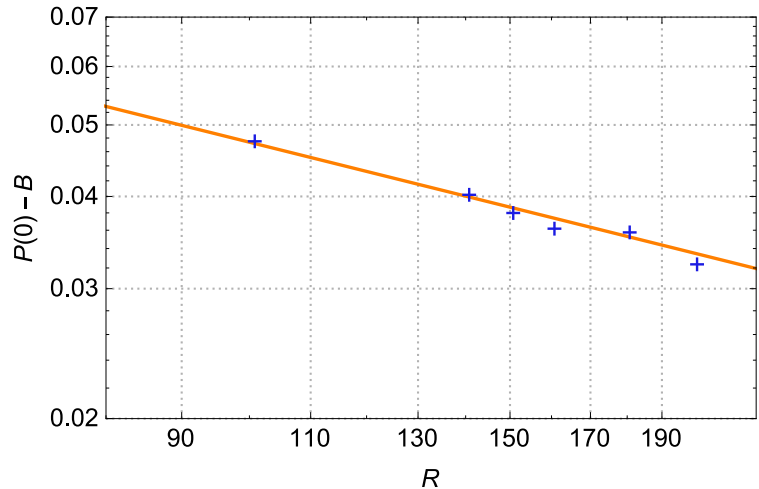}
\caption{Analytic result (\ref{p3}) for ${\cal P}(0)-B$ (solid line) and Monte Carlo data points at $T=0.985\,T_c$. The $R^{-1/2}$ suppression is the signature of the presence of a single interface (no interfacial wetting).
}
\label{figure04}
\end{figure}

We also show in Fig.~\ref{figure04} the comparison between theory and data for ${\cal P}(0)-B$. The logarithmic plot exhibits in particular the $R^{-1/2}$ suppression expected in absence of interfacial wetting. It is also clear that the correction of order $1/R$ is negligible for the temperature and lattice sizes considered. 

\section{Interfacial wetting}
As anticipated in the introduction, the $q$-state Potts model also allows to test the theoretical predictions about interfacial wetting. Indeed, for $q>4$ and $T=T_c$ two interfaces enclosing a macroscopic layer of the disorderd phase are expected to form in the interfacial region. The correlation length on the square lattice is exactly determined by \cite{BW} 
\EQ
\xi^{-1}=4 \sum_{n=0}^\infty\ln\left(\frac{1+[\sqrt{2}\cosh((\pi^2/2v)(n+\frac{1}{2}))]^{-1}}{1-[\sqrt{2}\cosh((\pi^2/2v)(n+\frac{1}{2}))]^{-1}}\right)\,,
\label{bw}
\EN
\EQ
2\cosh v=\sqrt{2+\sqrt{q}}\,,
\EN
and the quantitative predictions of the theory are expected to hold as long as $\xi$ is much larger than the lattice spacing \cite{bubbles}. In particular, the field theory result for the non-boundary color profile (\ref{profile}) is \cite{bubbles}
\begin{equation}
\label{dq}
\mathcal{P}(x) = \frac{q-2}{q} + \left(B -\frac{q-2}{q} \right)\, \frac{1+\mathcal{G}(\eta)}{2} + \cdots \,,
\end{equation}
where
\EQ
\eta=\frac{x}{\sqrt{R\xi}}\,,
\EN
\begin{equation}
\mathcal{G}(\eta) = -\frac{2}{\pi}\, \textrm{e}^{-2\eta^{2}} - \frac{2}{\sqrt{\pi}}\, \eta \, \textrm{erf}(\eta) \textrm{e}^{-\eta^{2}} + \textrm{erf}^{2}(\eta)\,,
\label{G}
\end{equation}
\EQ
\textrm{erf}(\eta)=\frac{2}{\sqrt{\pi}}\int_0^\eta \textrm{d}u\,e^{-u^2}\,,
\EN
and the corrections omitted in (\ref{dq}) are\footnote{It is a general result of the field theory of phase separation that corrections to the large $R$ asymptotics appear in powers of $R^{-1/2}$.} $O(R^{-1/2})$ at $x=0$. We see in particular the main signature of interfacial wetting, namely that ${\cal P}(0)$ approaches a constant as $R$ grows. It can be deduced from (\ref{dq}) that the probability $p_2(x_1,x_2)\textrm{d}x_1\textrm{d}x_2$ that the two interfaces enclosing the disordered phase intersect the axis $y=0$ in the intervals $(x_1,x_1+\textrm{d}x_1)$ and $(x_2,x_2+\textrm{d}x_2)$ is determined by
\begin{equation}
p_{2}(x_{1},x_{2}) = \frac{(\eta_{1}-\eta_{2})^{2}}{\pi R\xi}\,\textrm{e}^{-\eta_{1}^{2}-\eta_{2}^{2}}\,;
\label{passage2}
\end{equation}
notice the repulsion factor $(\eta_{1}-\eta_{2})^{2}$ that effectively makes the interfaces mutually avoiding \cite{bubbles,wedge2}. 

\begin{figure}
\centering
\includegraphics[width=10cm]{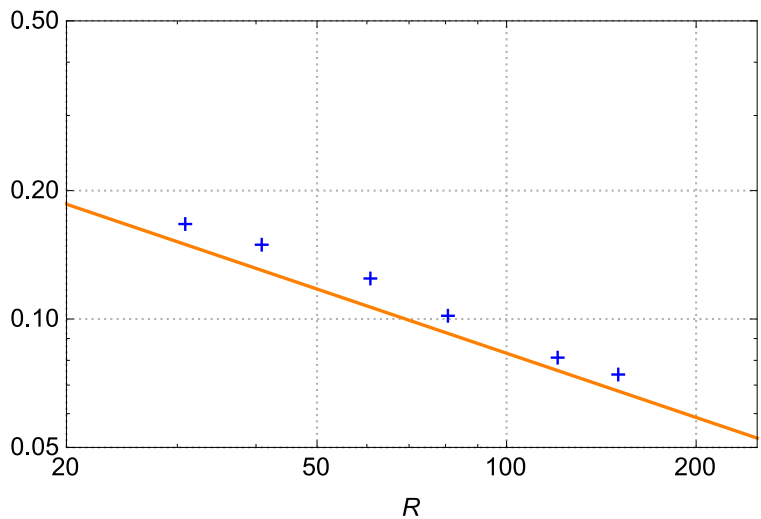}
\caption{The difference ${\cal P}(0)|_{R=\infty}-{\cal P}_{\textrm{num}}(0)$ for $q=10$ at $T=T_c$ and different values of $R$ (Monte Carlo data points). The solid line with slope ${-1/2}$ is a guide for the eye. Horizontal sizes up to $L=252$ have been used in the simulations.
}
\label{figure07}
\end{figure}

We determined ${\cal P}(x)$ by Monte Carlo simulations for $q=10$, a value for which (\ref{bw}) gives $\xi=10.55..$. In this case the first correction omitted in (\ref{dq}) is of order $R^{-1/2}$ and is still significant for $R=151$, the largest value of $R$ we simulated. Nonetheless, we can perform a non-trivial test of the analytic result (\ref{dq}) verifying whether the difference between the asymptotic value\footnote{We numerically determine $B\simeq 0.114$.} ${\cal P}(0)|_{R=\infty}=\frac{4}{5}+\frac{1}{2}\left(B-\frac{4}{5}\right)\left(1-\frac{2}{\pi}\right)\simeq 0.675$ and the numerically determined value ${\cal P}_{\textrm{num}}(0)$ is suppressed for large $R$ with the expected power $R^{-1/2}$. Figure~\ref{figure07} shows that this is indeed the case. The correction of order $R^{-1/2}$ in (\ref{dq}) can also be calculated within the formalism of \cite{bubbles}, but this goes beyond the scope of this paper.

Quite interestingly, we numerically verified that the main signatures for the excess probability ${\cal P}(0)-B$ persist also in regimes in which the constraint $\xi\gg a$ is not really satisfied. For example, we observed the $R^{-1/2}$ suppression at $T=0.9\,T_c$, as well as the approach to a constant at $T=T_c$ for $q=16$, a value of $q$ for which $\xi$ does not exceed four lattice spacings. 

\section{Conclusions}
In this paper we successfully tested against Monte Carlo simulations the predictions of the exact field theory of phase separation for planar systems at phase coexistence introduced in \cite{DV,bubbles}. We showed how the $q$-state Potts model for different ranges of temperature and $q$ allows the study of both the possible structures of the interfacial region: single interface and interfacial wetting. It is important to stress how these two regimes can be distinguished through local measurements of order parameter components; in principle, the interface passage probabilities (\ref{pd}) and (\ref{passage2}) can also be measured on the lattice, but this is much more demanding from the numerical point of view. A probability similar to (\ref{pd}) for $q=3$ below $T_c$ has been numerically investigated in \cite{Picco}.

The work of this paper can be extended in several directions. In the first place the generality of the theory calls for verification of its predictions in other models. For example, interfacial wetting is expected along the first order portion of the ferromagnetic-paramagnetic phase boundary in the Blume-Capel model \cite{bubbles,localization} (see \cite{FS} for a Monte Carlo study of interfacial adsorption in this model). The Ashkin-Teller model, on the other hand, allows for a continuously varying critical exponent $\nu$, and then for a transition between the single interface and interfacial wetting regimes \cite{bubbles,localization}. The present theory being formulated for pure systems, the interest of testing numerically interfacial wetting in presence of quenched disorder was pointed out in \cite{potts_random}. Finally, it should be mentioned that the field theory for boundary conditions producing topological defects (interfaces in two dimensions) has been extended to higher dimensions in \cite{topological}. In particular, in that paper an order parameter profile has been exactly predicted for the $XY$ model in three dimensions and its numerical confirmation would be theoretically relevant.

\vspace{1cm} \noindent \textbf{Acknowledgments.} A.S. thanks the Galileo Galilei Institute for Theoretical Physics in Arcetri for hospitality during the winter school \emph{``SFT 2018: Lectures on Statistical Field Theories''}, where part of this work was carried out.


\end{document}